\documentclass[12pt]{article}
\pdfoutput=1

\usepackage{amsmath}
\usepackage{amssymb}
\usepackage{epsfig}
\usepackage{epstopdf}
\usepackage{latexsym}
\usepackage{color}
\numberwithin{equation}{section}
\usepackage[
      colorlinks=false,
      linkcolor=darkblue,  
      urlcolor=blue,    
      filecolor=blue,     
      citecolor=red,
linktocpage=true,
      pdfstartview=FitV,
      bookmarksopen=true    
      ]{hyperref}

\begin{document}

\begin{titlepage}

\centerline{\huge \rm } 
\bigskip
\centerline{\huge \rm Holographic renormalization group flows}
\bigskip
\centerline{\huge \rm in two-dimensional gravity and $AdS$ black holes}
\bigskip
\bigskip
\bigskip
\bigskip
\bigskip
\bigskip
\centerline{\rm Minwoo Suh}
\bigskip
\centerline{\it Department of Physics, Kyungpook National University, Daegu 41566, Korea}
\bigskip
\centerline{\tt minwoosuh1@gmail.com} 
\bigskip
\bigskip
\bigskip
\bigskip
\bigskip
\bigskip
\bigskip
\bigskip
\bigskip
\bigskip

\begin{abstract}
\noindent We look into the $AdS$ black holes from two-dimensional gravity perspective. In this work, we extend the previous results of holographic renormalization group flows to dimensions two. By introducing a superpotential, we derive the flow equations in two-dimensional dilaton gravity. We find a quantity which monotonically decreases along flows and give some comments on holographic $c$-theorem. As examples, we show that recently studied supersymmetric $AdS$ black hole solutions generically dimensionally reduce to two-dimensional dilaton gravity, and obtain the flow equations for black hole solutions.
\end{abstract}

\vskip 5cm

\flushleft {February, 2020}

\end{titlepage}

\tableofcontents

\section{Introduction}

Recently, via the AdS/CFT correspondence, \cite{Maldacena:1997re}, following the grounding works of \cite{Cacciatori:2009iz, Benini:2015noa, Benini:2015eyy}, there has been remarkable development in microscopic counting of microstates of supersymmetric $AdS$ black holes via twisted or superconformal indices of dual field theories. It was shown that entropy functional extremizes to give the Bekenstein-Hawking entropy of the black holes. In the dual field theory, this extremization procedure is understood as the $\mathcal{I}$-extremization where the Witten index of $1d$ superconformal quantum mechanics is extremized to give the entropy. $1d$ quantum mechanics is dual to the $AdS_2$ horizon geometry of extremal black hole solutions. See \cite{Zaffaroni:2019dhb} for a review and references.

Therefore, it is natural to try to understand the $\mathcal{I}$-extremization principle from two-dimensional supergravity point of view. Indeed, $a$-maximization of $4d$ SCFTs \cite{Intriligator:2003jj}, $c$-extremization of $2d$ SCFTs, \cite{Benini:2012cz, Benini:2013cda}, and $\tau_{RR}$-extremization of $3d$ SCFTs, \cite{Barnes:2005bm}, were understood from five-\cite{Tachikawa:2005tq, Szepietowski:2012tb}, three- \cite{Karndumri:2013iqa, Karndumri:2013dca, Karndumri:2015sia} and four-\cite{Amariti:2015ybz} dimensional gauged supergravity theories by identifying the corresponding quantities of $a$, $c$, and $\tau_{RR}$, respectively. On the other hand, even though supergravity in two dimensions has been studied long enough, \cite{Nicolai:1998gi, Paulot:2006zp, Samtleben:2007an, Ortiz:2012ib, Anabalon:2013zka, Bossard:2018utw}, our understanding is elementary and two-dimensional supergravity models are rare compare to the other dimensional theories.

For domain wall backgrounds of gravity theories in dimensions higher than two, by introduction of superpotential, the second order equations of motion often reduce to first order flow equations, \cite{Freedman:1999gp, Skenderis:1999mm, DeWolfe:1999cp}. The flow equations arises naturally, as they are the Hamilton-Jacobi equations of gravity-scalar theories, \cite{deBoer:1999tgo}. In supergravity, the flow equations, in fact, reproduces the BPS equations obtained from the supersymmetry variations, \cite{Freedman:1999gp}. They also provide the holographic $c$-theorem, \cite{Freedman:1999gp, Skenderis:1999mm, Girardello:1998pd}. Solutions of the flow equations are known to be non-perturbatively stable, \cite{Freedman:2003ax}. See section 9 and 10 of \cite{DHoker:2002nbb} for a review. However, the flow equations are known in dimensions higher than two. 

In this work, we extend the known results of holographic renormalization group flows in dimensions higher than two and derive the flow equations in two-dimensional dilaton gravity. We begin by considering general two-dimensional dilaton gravity coupled to a scalar field. By introducing a superpotential, we derive the first order flow equations from the equations of motion. We find a quantity which decreases monotonically along flows and give some comments on holographic $c$-theorem.

As examples, we show that recently studied supersymmetric $AdS$ black holes generically dimensionally reduce to two-dimensional dilaton gravity. To be specific, we consider the supersymmetric black holes in $AdS_4$, \cite{Cacciatori:2009iz}, $AdS_6$, \cite{Suh:2018tul}, and $AdS_5$, \cite{Nieder:2000kc}. Their microstates are counted by topologically twisted index of $3d$, \cite{Benini:2015noa, Benini:2015eyy}, $5d$, \cite{Hosseini:2018uzp, Crichigno:2018adf}, and, $4d$, \cite{Bae:2019poj}, dual field theories, respectively. We present the flow equations of higher-dimensional $AdS$ black holes in two dimensions. We also show that from the dilaton at the $AdS_2$ fixed point, we can obtain the Bekenstein-Hawking entropy of higher-dimensional $AdS$ black holes, $e.g.$, \cite{Myers:1994sg} and \cite{Gegenberg:1994pv}.

In section 2, we review the flow equations of gravity in dimensions higher than two. In section 3, we consider two-dimensional dilaton gravity and derive their flow equations. We find a quantity which monotonically decrease along flows and give some comments on holographic $c$-theorem. In section 4, as examples, we obtain two-dimensional dilaton gravity from supersymmetric $AdS_4$, $AdS_6$, and $AdS_5$ black holes. We present their flow equations. We conclude in section 5.

\section{Review of gravity in dimensions higher than two}

For domain wall backgrounds in dimensions higher than two, the second order equations of motion reduce to first order flow equations, \cite{Freedman:1999gp, Skenderis:1999mm, DeWolfe:1999cp}. The flow equations arise naturally, as they are the Hamilton-Jacobi equations of dynamical system of gravity-scalar theories, \cite{deBoer:1999tgo}. In supergravity, the first order equations reproduce the BPS equations obtained from the supersymmetry variations, \cite{Freedman:1999gp}. Solutions of the equations are known to be non-perturbatively stable, \cite{Freedman:2003ax}. In this section, we review section 9 of \cite{DHoker:2002nbb}. We consider the gravity coupled to a scalar field,
\begin{equation}
S\,=\,\frac{1}{4\pi{G}_N^{(d+1)}}\int{d}^{d+1}x\sqrt{+g}\left[-\frac{1}{4}R+\frac{1}{2}\partial_\mu\phi\partial^\mu\phi+V\right]\,,
\end{equation}
in Euclidean spacetime. The equations of motion are
\begin{align}
R_{\mu\nu}-\frac{1}{2}Rg_{\mu\nu}\,&=\,2\left[\partial_\mu\phi\partial_\nu\phi-g_{\mu\nu}\left(\frac{1}{2}\partial_\rho\phi\partial^\rho\phi+V\right)\right]\,, \notag \\
\frac{1}{\sqrt{g}}\partial_\mu\left(\sqrt{g}g^{\mu\nu}\partial_\nu\phi\right)-\frac{dV}{d\phi}\,&=\,0\,.
\end{align}
Therefore, at critical points, the scalar potential satisfies
\begin{equation} \label{fpoint}
\left.\frac{dV}{d\phi}\right|_*\,=\,0\,.
\end{equation}

\subsection{Flow equations}

We consider the domain wall background,
\begin{equation} \label{flatdw}
ds^2\,=\,e^{2A}ds_{d}^2+dr^2\,.
\end{equation}
The equations of motion reduce to
\begin{align} \label{eqeq}
(d-1)A''+\frac{d(d-1)}{2}A'^2\,=&\,-\phi'^2-2V\,, \notag \\ 
\frac{d(d-1)}{2}A'^2\,=&\,\phi'^2-2V\,, \notag \\ 
\phi''+dA'\phi'\,=&\,\frac{dV}{d\phi}\,,
\end{align}
where the primes denote the derivative with respect to $r$. However, the first equation is obtained from the last two equations. Hence, there are only two independent equations of motion,
\begin{align}
A'^2\,=&\,\frac{2}{d(d-1)}\left(\phi'^2-2V\right)\,, \notag \\
\phi''+dA'\phi'\,=&\,\frac{dV}{d\phi}\,.
\end{align}
Miraculously, by employing a superpotential, $W$, the second order equations reduce to first order flow equations,
\begin{align} \label{firstordersystem}
\frac{d\phi}{dr}\,=&\,\frac{dW}{d\phi}\,, \notag \\
\frac{dA}{dr}\,=&\,-\frac{2}{d-1}W\,,
\end{align}
where the superpotential generates the scalar potential,
\begin{equation}
V\,=\,\frac{1}{2}\left(\frac{dW}{d\phi}\right)^2-\frac{d}{d-1}W^2\,.
\end{equation}
This result trivially extends to theories with multiple scalar fields. A large class of solutions in supergravity falls into the flat domain wall backgrounds we considered, like, holographic RG flows, and easily extended to wrapped branes, and black hole solutions. For more complicated backgrounds, like Janus solutions, \cite{Freedman:2003ax, Clark:2005te, Suh:2011xc, Bobev:2013yra}, which are $AdS$ domain walls, and also sphere domain walls, the basic structure of the flow equations stays the same, but gets more involved. 

\subsection{Holographic $c$-theorem}

The, so-called, monotonicity theorems provide measures of degrees of freedom which monotonically decreases along the renormalization group flows in field theories. There are $c$-theorem, \cite{Zamolodchikov:1986gt}, $a$-theorem, \cite{Cardy:1988cwa, Komargodski:2011vj}, and $F$-theorem, \cite{Jafferis:2011zi, Casini:2012ei}, in two-, four-, and three-dimensional field theories, respectively. There are attempts at $a$-theorem in six-dimensional field theories, $e.g$, \cite{Elvang:2012st, Grinstein:2014xba, Cordova:2015fha}. 

One of the immediate applications of holographic RG flows is the holographic $c$-theorem, \cite{Girardello:1998pd, Freedman:1999gp, Skenderis:1999mm} and \cite{Myers:2010xs, Myers:2010tj}. See section 10 of \cite{DHoker:2002nbb} for a review. Though the following argument is valid in $d>1$ dimensions, we will restrict to the AdS$_5$/CFT$_4$ correspondence as it provides a holographic $c$-function which is indeed dual to the $c$-function of four-dimensional field theories, \cite{Freedman:1999gp}. For the domain wall background, from the difference of the first and the second equations in \eqref{eqeq}, we find that
\begin{equation} \label{pos}
A''\,=\,-\frac{2}{d-1}\phi'^2\,\leq\,0.
\end{equation}
We define a function,
\begin{equation}
C(r)\,=\,\frac{\pi}{8G_N}\frac{1}{A'^3}\,.
\end{equation}
Due to \eqref{pos}, derivative of the function is always nonnegative,
\begin{equation}
C'(r)\,=\,\frac{\pi}{8G_N}\frac{-3A''}{A'^4}\,\geq\,0\,.
\end{equation}
Therefore, this function naturally introduces a holographic $c$-function. Moreover, from the flow equations, \eqref{firstordersystem}, we have, \cite{Freedman:1999gp},
\begin{equation}
C(r)\,\sim\,-\frac{1}{W^3}\,.
\end{equation}
See \cite{Szepietowski:2012tb} also for a relation of holographic $a$-theorem and $a$-maximization.

\section{Two-dimensional dilaton gravity}

We consider two-dimensional dilaton gravity coupled to a scalar field,
\begin{equation}
S\,=\,\frac{1}{16\pi{G}_N^{(2)}}\int{d}^2x\sqrt{-g}e^{2\Phi}\left[R+2(2\alpha+1)\partial_\mu\Phi\partial^\mu\Phi-\frac{1}{2}\partial_\mu\phi\partial^\mu\phi-e^{-2\Phi}V(\Phi, \phi; \alpha)\right]\,,
\end{equation}
where $\Phi$ is the dilaton and $\phi$ is the scalar field. In two dimensions, we cannot go to Einstein frame by performing conformal transformations, and they just transform to other string frames. Choosing value of the constant parameter, $\alpha$, is equivalent of conformal transformations. When we study examples from higher dimensions latter in \eqref{alphaa}, it will be clear why we have chosen the factor to be $(2\alpha+1)$. See \cite{Grumiller:2002nm} for a review of two-dimensional dilaton gravity.

The equations of motion are
\begin{align} \label{eomtwo}
R_{\mu\nu}-\frac{1}{2}Rg_{\mu\nu}-e^{-2\Phi}\left(\nabla_\mu\nabla_\nu{e}^{2\Phi}-g_{\mu\nu}\nabla_\rho\nabla^\rho{e}^{2\Phi}\right) \,\,\,\,\,\,\,\,\,\,\,\,\,\,\,\,\,\,\,\,\,\,\,\,\,\,\,\,\,\,\,\,\,\,\,\,\,\,\,\,\,\,\,\,\,\,\,\,\,\,\,\,\,\,\,\,\,\,\,\,\,\,\,\,\,\,\,\,\,\,\,\,\,\,\,\,\,\,\,\,\,\, \notag \\ 
=\,-2(2\alpha+1)\partial_\mu\Phi\partial_\nu\Phi+(2\alpha+1)g_{\mu\nu}\partial_\rho\Phi\partial^\rho\Phi+\frac{1}{2}\partial_\mu\phi\partial_\nu\phi-\frac{1}{4}g_{\mu\nu}\partial_\rho\phi\partial^\rho\phi-\frac{1}{2}e^{-2\Phi}Vg_{\mu\nu}&\,, \notag \\
2(2\alpha+1)\frac{1}{\sqrt{-g}}\partial_\mu\left(\sqrt{-g}g^{\mu\nu}\partial_\nu\Phi\right)-2(2\alpha+1)\partial_\mu\Phi\partial^\mu\Phi-\frac{1}{2}\partial_\mu\phi\partial^\mu\phi+R-\frac{1}{2}e^{-2\Phi}\frac{\partial{V}}{\partial\Phi}\,=&\,0\,, \notag \\ 
\frac{1}{\sqrt{-g}}\partial_\mu\left(\sqrt{-g}g^{\mu\nu}\partial_\nu\phi\right)+2\partial_\mu\Phi\partial^\mu\phi-e^{-2\Phi}\frac{\partial{V}}{\partial\phi}\,=&\,0\,.
\end{align}
On the left hand side of the Einstein equation in \eqref{eomtwo}, in addition to the Einstein tensor, there are derivative terms of the dilaton. In two dimensions, the Einstein tensor vanishes identically,
\begin{equation}
R_{\mu\nu}-\frac{1}{2}Rg_{\mu\nu}\,=\,0\,.
\end{equation}

There are two classes of solutions: i) solutions with constant dilaton and scalar field and ii) solutions with linear dilaton. See, for instance, around (12) and (13) of \cite{Grumiller:2014oha}. The first class of solutions are maximally symmetric, $i.e.$, Minkowski, de Sitter, and anti-de Sitter, and we will concentrate on the first class of solutions. For the first class of solutions, $\Phi\,=\,\Phi_*$ and $\phi\,=\,\phi_*$, from the equations of motion, we obtain
\begin{equation} \label{fpoint}
V|_*\,=\,0\,, \qquad R-\left.\frac{1}{2}e^{-2\Phi}\frac{\partial{V}}{\partial\Phi}\right|_*\,=\,0\,, \qquad \left.\frac{\partial{V}}{\partial\phi}\right|_*\,=\,0\,.
\end{equation}
Curvatures of the solutions are determined by the second equation in \eqref{fpoint}. For the solutions satisfying the conditions, $i.e.$, constant dilaton and scalar field with maximally symmetric background, we refer to them as critical points.

\subsection{Flow equations}

We consider the domain wall background,
\begin{equation} \label{background}
ds^2\,=\,e^{2A}\left(-dt^2+dr^2\right)\,.
\end{equation}
The equations of motion reduce to
\begin{align} \label{fos}
2\left(\Phi''+2\Phi'\Phi'\right)-2A'\Phi'-(2\alpha+1)\Phi'\Phi'+\frac{1}{4}\phi'\phi'+\frac{1}{2}e^{-2\Phi+2A}V\,&=\,0\,, \notag \\ 
-2A'\Phi'-(2\alpha+1)\Phi'\Phi'+\frac{1}{4}\phi'\phi'-\frac{1}{2}e^{-2\Phi+2A}V\,&=\,0\,, \notag \\ 
(2\alpha+1)\left(\Phi''+\Phi'\Phi'\right)+A''+\frac{1}{4}\phi'\phi'+\frac{1}{4}e^{-2\Phi+2A}\frac{\partial{V}}{\partial\Phi}\,&=\,0\,, \notag \\ 
\phi''+2\Phi'\phi'-e^{-2\Phi+2A}\frac{\partial{V}}{\partial\phi}\,&=\,0\,,
\end{align}
where the primes denote the derivative with respect to $r$.

From the sum of the first and the second equations in \eqref{fos}, we obtain a relation between derivatives of functions without the scalar potential,
\begin{equation} \label{eomdifference}
\left(\Phi''+2\Phi'\Phi'\right)-2A'\Phi'-(2\alpha+1)\Phi'\Phi'+\frac{1}{4}\phi'\phi'\,=\,0\,.
\end{equation}
Starting from this relation, we look for the first order flow equations by trial and error. By introducing a superpotential, $W$, we obtain the flow equations, 
\begin{align} \label{2flow}
\frac{d\Phi}{dr}e^{-A}\,=&\,W\,, \notag \\
\frac{d\phi}{dr}e^{-A}\,=&\,-4\frac{\partial{W}}{\partial\phi}\,, \notag \\
\frac{dA}{dr}e^{-A}\,=&\,\frac{\partial{W}}{\partial\Phi}+2W-(2\alpha+1)W\,.
\end{align}
Therefore, at critical points, the superpotential satisfies
\begin{equation} \label{fpoint2}
W|_*\,=\,0\,, \qquad \left.\frac{\partial{W}}{\partial\phi}\right|_*\,=\,0\,.
\end{equation}
The superpotential produces the scalar potential by
\begin{equation}
V\,=\,e^{2\Phi}\left[8\left(\frac{\partial{W}}{\partial\phi}\right)^2-4W\frac{\partial{W}}{\partial\Phi}-8W^2+2(2\alpha+1)W^2\right]\,.
\end{equation}
This result trivially extends to theories with multiple scalar fields.{\footnote {Similar first order equations in two-dimensional dilaton gravity were also considered in \cite{Castro:2018ffi}, see (5.14) and appendix A.1 therein.}}

\subsection{Comments on holographic $c$-theorem}

In this subsection, we analogously follow the derivation of holographic $c$-theorem in dimensions higher than two in section 2.2. From the difference of two equations of motion, \eqref{eomdifference}, we obtain
\begin{equation}
\left(e^{-2A}\Phi'\right)'\,=\,\left((2\alpha+1)-2\right)e^{-2A}\Phi'\Phi'-\frac{1}{4}e^{-2A}\phi'\phi'\,\leq\,0\,,
\end{equation}
where we choose the free parameter, $\alpha$, to be{\footnote{At $\alpha\,=\,1/2$, the kinetic term of the dilaton field is canonically normalized. As we can freely perform conformal transformations, we can choose $\alpha$ to be at any value.}}
\begin{equation} \label{alhalf}
\alpha\,\leq\,\frac{1}{2}\,.
\end{equation}
From the supersymmetry equations, \eqref{2flow}, we also find that
\begin{equation}
e^{-2A}\Phi'\,=\,e^{-A}W\,.
\end{equation}
Thus, finally, we find a quantity,
\begin{equation}
C(r)\,=\,e^{-A}W\,,
\end{equation}
whose derivative is negative or zero,
\begin{equation} \label{md}
C'(r)\,\leq\,0\,.
\end{equation}

In fact, as $C(r)$ vanishes at critical points by \eqref{fpoint2}, this quantity may not play the role of holographic $c$-function. However, whether it is a holographic $c$-function or not, there is an intriguing interpretation of the quantity. From \eqref{fpoint2} and \eqref{md}, we have
\begin{align} \label{cd1}
W|_*\,=\,0\,, \\  \label{cd2}
(e^{-A}W)'\,\leq\,0\,.
\end{align}
These conditions mean that $C(r)$ vanishes at critical points, \eqref{cd1}, and $C(r)$ decreases monotonically along the flows, \eqref{cd2}. Therefore, it implies that there can be $only$ one critical point. 

In the following section, we will see that the only critical point corresponds to the IR fixed point from higher dimensions, which is the near horizon in the case of black hole solutions. However, there is no UV fixed point in two-dimensional gravity. The running dilaton in the UV means that the two-dimensional description breaks down and one must uplift to higher dimensions.{\footnote{We would like to thank the anonymous referee for comment on this.}} An example of such a flow is discussed in section 2.3 of \cite{Cvetic:2016eiv}, in which case, the UV fixed point is the BTZ black hole in $AdS_3$.

\section{Two-dimensional dilaton gravity from gauged supergravity}

\subsection{Supersymmetric $AdS_4$ black holes}

We review the supersymmetric $AdS_4$ black hole solutions of \cite{Cacciatori:2009iz, DallAgata:2010ejj, Hristov:2010ri} from gauged $\mathcal{N}\,=\,2$ supergravity. We employ the conventions of appendix A in \cite{Benini:2015eyy}. Their microstates are counted by topologically twisted index of $3d$ SCFTs, \cite{Benini:2015noa, Benini:2015eyy}. The action is given by{\footnote {In \cite{Benini:2015eyy} the scalar fields are denoted by $\phi_1\,=\,\phi_{12}$, $\phi_2\,=\,\phi_{13}$, $\phi_3\,=\,\phi_{14}$.}}
\begin{equation}
S\,=\,\frac{1}{16\pi{G}_N^{(4)}}\int{d^4x}\sqrt{-g_4}\left[R_4-\sum_{i=1}^3\frac{1}{2}\partial_\mu\phi_i\partial^\mu\phi_i-\sum_{a=1}^4\frac{1}{2}L_a^{-2}F_a^2-V_4\right]\,,
\end{equation}
where the scalar potential is
\begin{align}
V_4\,=&\,-4\tilde{g}^2\left(\cosh\phi_1+\cosh\phi_2+\cosh\phi_3\right)\,, \notag \\
=&\,-2\tilde{g}^2\left(L_1L_2+L_3L_4+L_1L_3+L_2L_4+L_1L_4+L_2L_3\right)\,,
\end{align}
If we define the superpotential by
\begin{equation}
W_4=\,-\frac{1}{4}\left(L_1+L_2+L_3+L_4\right)\,,
\end{equation}
then, the scalar potential is obtained from
\begin{equation}
V_4\,=\,8\sum_{i=1}^3\left(\frac{\partial{W_4}}{\partial\phi_i}\right)^2-6W_4^2\,.
\end{equation}
We set the gauge coupling constant to be
\begin{equation}
\tilde{g}\,=\,\frac{1}{\sqrt{2}}\,.
\end{equation}
We introduced a parametrization of three real scalar fields,
\begin{align}
L_1\,&=\,e^{-\frac{1}{2}\left(\phi_1+\phi_2+\phi_3\right)}\,, \qquad L_2\,=\,e^{-\frac{1}{2}\left(\phi_1-\phi_2-\phi_3\right)}\,, \notag \\
L_3\,&=\,e^{-\frac{1}{2}\left(-\phi_1+\phi_2-\phi_3\right)}\,, \qquad L_4\,=\,e^{-\frac{1}{2}\left(-\phi_1-\phi_2+\phi_3\right)}\,,
\end{align}
where $L_1L_2L_3L_4\,=\,1$. The field strength of four $U(1)$ gauge fields are
\begin{equation}
F_a\,=\,dA_a\,.
\end{equation}

For the supersymmetric $AdS_4$ black hole solutions of \cite{Cacciatori:2009iz}, we consider the background of
\begin{equation}
ds^2\,=\,e^{2f}\left(-dt^2+dr^2\right)+e^{2g}ds^2_\Sigma\,,
\end{equation}
where $\Sigma$ denotes the Riemann surfaces of curvatures, $k\,=\,\pm1$. The field strength of the gauge fields are 
\begin{equation}
F_a\,=\,-\frac{a_a}{\sqrt{2}}Vol_\Sigma\,,
\end{equation}
where the magnetic charges, $a_a$, are constant and $Vol_\Sigma$ is the unit volume form. The first order BPS equations are obtained by solving the supersymmetry variations of the fermionic fields,{\footnote{We suspect that the signs in front of $e^{-2g}$ should be flipped in (A.27) of \cite{Benini:2015eyy}.}}
\begin{align} \label{ads4susy}
f'e^{-f}\,=\,-\frac{1}{4}\left(L_1+L_2+L_3+L_4\right)-\frac{1}{4}e^{-2g}\left(\frac{a_1}{L_1}+\frac{a_2}{L_2}+\frac{a_3}{L_3}+\frac{a_4}{L_4}\right)\,, \notag \\
g'e^{-f}\,=\,-\frac{1}{4}\left(L_1+L_2+L_3+L_4\right)+\frac{1}{4}e^{-2g}\left(\frac{a_1}{L_1}+\frac{a_2}{L_2}+\frac{a_3}{L_3}+\frac{a_4}{L_4}\right)\,, \notag \\
\phi_1'e^{-f}\,=\,-\frac{1}{2}\left(L_1+L_2-L_3-L_4\right)-\frac{1}{2}e^{-2g}\left(\frac{a_1}{L_1}+\frac{a_2}{L_2}-\frac{a_3}{L_3}-\frac{a_4}{L_4}\right)\,, \notag \\
\phi_2'e^{-f}\,=\,-\frac{1}{2}\left(L_1-L_2+L_3-L_4\right)-\frac{1}{2}e^{-2g}\left(\frac{a_1}{L_1}-\frac{a_2}{L_2}+\frac{a_3}{L_3}-\frac{a_4}{L_4}\right)\,, \notag \\
\phi_3'e^{-f}\,=\,-\frac{1}{2}\left(L_1-L_2-L_3+L_4\right)-\frac{1}{2}e^{-2g}\left(\frac{a_1}{L_1}-\frac{a_2}{L_2}-\frac{a_3}{L_3}+\frac{a_4}{L_4}\right)\,,
\end{align}
and there is a twist condition on the magnetic charges,
\begin{equation}
a_1+a_2+a_3+a_4\,=\,\frac{k}{\tilde{g}^2}\,.
\end{equation}
One can solve the BPS equations and obtain the supersymmetric black hole solutions which are interpolating between the $AdS_4$ boundary and the $AdS_2$ horizon.

There is an $AdS_2\times\Sigma$ solution of the horizon. We introduce another parametrization of the scalar fields,
\begin{equation}
z_{1,2,3}\,=\,\frac{L_{1,2,3}}{L_4}\,,
\end{equation}
or, equivalently,
\begin{equation}
L_1^4\,=\,\frac{z_1^3}{z_2z_3}\,, \qquad L_2^4\,=\,\frac{z_2^3}{z_3z_1}\,, \qquad L_3^4\,=\,\frac{z_3^3}{z_1z_2}\,, \qquad L_4^4\,=\,\frac{1}{z_1z_2z_3}\,.
\end{equation}
The horizon is at
\begin{align} \label{ads4ads211}
z_1\,=\,\frac{2(a_2+a_3)(a_1-a_4)^2-(a_1+a_4)[(a_2-a_3)^2+(a_1-a_4)^2]+4(a_4-a_1)\sqrt{\Theta}}{2a_4(a_4-a_1+a_2-a_3)(a_4-a_1-a_2+a_3)}\,, \notag \\
z_2\,=\,\frac{2(a_1+a_3)(a_2-a_4)^2-(a_2+a_4)[(a_1-a_3)^2+(a_2-a_4)^2]+4(a_4-a_2)\sqrt{\Theta}}{2a_4(a_4+a_1-a_2-a_3)(a_4-a_1-a_2+a_3)}\,, \notag \\
z_3\,=\,\frac{2(a_1+a_2)(a_3-a_4)^2-(a_3+a_4)[(a_1-a_2)^2+(a_3-a_4)^2]+4(a_4-a_3)\sqrt{\Theta}}{2a_4(a_4+a_1-a_2-a_3)(a_4-a_1+a_2+a_3)}\,,
\end{align}
and
\begin{align} \label{ads4ads222}
e^{2f}\,=&\,\frac{1}{r^2}\frac{\Pi}{\sqrt{2}\Theta}\left(F_2+\sqrt{\Theta}\right)^{1/2}\,, \notag \\
e^{2g}\,=&\,\frac{1}{\sqrt{2}}\left(F_2+\sqrt{\Theta}\right)^{1/2}\,,
\end{align}
where
\begin{align}
\Theta\,=&\,F_2^2-4a_1a_2a_3a_4\,, \notag \\
F_2\,=&\,\frac{1}{4}\left(a_1+a_2+a_3+a_4\right)^2-\frac{1}{2}\left(a_1^2+a_2^2+a_3^2+a_4^2\right)\,, \notag \\
\Pi\,=&\,\frac{1}{8}(a_1+a_2-a_3-a_4)(a_1-a_2+a_3-a_4)(a_1-a_2-a_3+a_4)\,.
\end{align}

Now we dimensionally reduce the action on the background of supersymmetric $AdS_4$ black holes. The reduction ansatz for the metric is 
\begin{equation}\label{alphaa}
ds^2_4\,=\,e^{2\alpha{g}}ds_2^2+e^{2g}ds^2_\Sigma\,,
\end{equation}
where $\alpha$ is a constant parameter. The reduced action is two-dimensional dilaton gravity,
\begin{equation}
S\,=\,\frac{vol_\Sigma}{16\pi{G}_N^{(4)}}\int{d^2x}\sqrt{-g_2}e^{2g}\left[R_2+2(2\alpha+1)\partial_\mu{g}\partial^\mu{g}-\sum_{i=1}^3\frac{1}{2}\partial_\mu\phi_i\partial^\mu\phi_i-e^{-2g}V_2\right]\,,
\end{equation}
where the scalar potential is
\begin{align}
V_2\,=\,e^{2(\alpha+1){g}}\Big[&-2\tilde{g}^2\left(L_1L_2+L_3L_4+L_1L_3+L_2L_4+L_1L_4+L_2L_3\right) \notag \\
&+e^{-4g}\left(\frac{a_1^2}{L_1^2}+\frac{a_2^2}{L_2^2}+\frac{a_3^2}{L_3^2}+\frac{a_4^2}{L_4^2}\right)-2ke^{-2g}\Big]\,.
\end{align}
The scalar potential satisfies the relations, \eqref{fpoint}, at the $AdS_2$, critical point in \eqref{ads4ads211} and \eqref{ads4ads222}.

We consider the background,
\begin{equation}
ds^2_2\,=\,e^{2f}\left(-dt^2+dr^2\right)\,.
\end{equation}
We obtain the flow equations which satisfy the equations of motion of two-dimensional dilaton gravity,
\begin{align}
f'e^{-f}\,=&\,W+\frac{\partial{W}}{\partial{g}}\,, \notag \\
g'e^{-f}\,=&\,W\,, \notag \\
\phi'_ie^{-f}\,=&\,-4\frac{\partial{W}}{\partial\phi_i}\,,
\end{align}
where the superpotential is given by
\begin{equation}
W=\,-\frac{1}{4}\left(L_1+L_2+L_3+L_4\right)+\frac{1}{4}e^{-2g}\left(\frac{a_1}{L_1}+\frac{a_2}{L_2}+\frac{a_3}{L_3}+\frac{a_4}{L_4}\right)\,.
\end{equation}
The superpotential satisfies the relations, \eqref{fpoint2}, at the $AdS_2$ critical point in \eqref{ads4ads211} and \eqref{ads4ads222}. The scalar potential is obtained from
\begin{equation}
V_2\,=\,e^{2g}\left[8\sum_{i=1}^3\left(\frac{\partial{W}}{\partial\phi_i}\right)^2-4W\frac{\partial{W}}{\partial{g}}-6W^2\right]\,.
\end{equation}
By introducing the superpotential, we note that the flow equations are merely a rewriting of the BPS equations in $AdS_4$, \eqref{ads4susy}, as it should. Moreover, we could reproduce the flow equations from the general analysis of section 2 by replacing
\begin{equation}
\Phi\,\rightarrow\,g\,, \qquad \phi\,\rightarrow\,\phi_i\,, \qquad A\,\rightarrow\,f\,, \qquad W\rightarrow\,W\,, \qquad V\,\rightarrow\,V_2\,,
\end{equation}
and setting
\begin{equation}
\alpha\,=\,0\,.
\end{equation}

Finally, we reproduce the Bekenstein-Hawking entropy of the supersymmetric $AdS_4$ black holes obtained in \cite{Benini:2015eyy}, 
\begin{equation}
S_{\text{BH}}\,=\,\left.\frac{e^{2g}vol_{\Sigma}}{4G_N^{(4)}}\right|_*\,,
\end{equation}
where $vol_\Sigma$ is the area of Riemann surfaces of unit radius, in this case. We make an observation that the two-dimensional dilaton, $\Phi\,\rightarrow\,g$, contains the information of the Bekenstein-Hawking entropy of higher dimensional $AdS$ black holes.

\subsection{Supersymmetric $AdS_6$ black holes}

We review the supersymmetric $AdS_6$ black hole solutions of pure $F(4)$ gauged supergravity in \cite{Suh:2018tul}, which were generalized to solutions of matter coupled $F(4)$ gauged supergravity in \cite{Hosseini:2018usu, Suh:2018szn}. Their microstates are microscopically counted by topologically twisted index of $5d$ SCFTs, \cite{Hosseini:2018uzp}. See also \cite{Crichigno:2018adf}.  

The bosonic field content of pure $F(4)$ gauged supergravity, \cite{Romans:1985tw}, consists of the metric, $g_{\mu\nu}$, a real scalar, $\phi$, an $SU(2)$ gauge field, $A^I_\mu$, $I\,=\,1,\,2,\,3$, a $U(1)$ gauge field, $\mathcal{A}_\mu$, and a two-form gauge potential, $B_{\mu\nu}$. The field strengths are defined by
\begin{align}
\mathcal{F}_{\mu\nu}\,=&\,\partial_\mu\mathcal{A}_\nu-\partial_\nu\mathcal{A}_\mu\,, \notag \\
F^I_{\mu\nu}\,=&\,\partial_\mu{A}^I_\nu-\partial_\nu{A}^I_\mu+\tilde{g}\epsilon^{IJK}A^J_\mu{A}^K_\nu\,, \notag \\
G_{\mu\nu\rho}\,=&\,3\partial_{[\mu}B_{\nu\rho]}\,, \notag \\
\mathcal{H}_{\mu\nu}\,=&\,\mathcal{F}_{\mu\nu}+mB_{\mu\nu}\,.
\end{align}
The action is given by
\begin{align}
S\,=\,\frac{1}{4\pi{G}_N^{(6)}}\int{d^6x}\sqrt{-g_6}&\left[-\frac{1}{4}R+\frac{1}{2}\partial_\mu\phi\partial^\mu\phi-\frac{1}{4}e^{-\sqrt{2}\phi}\left(\mathcal{H}_{\mu\nu}\mathcal{H}^{\mu\nu}+F^I_{\mu\nu}F^{I\mu\nu}\right)+\frac{1}{12}e^{2\sqrt{2}\phi}G_{\mu\nu\rho}G^{\mu\nu\rho}\right. \notag \\
&\left.-\frac{1}{8}\epsilon^{\mu\nu\rho\sigma\tau\kappa}B_{\mu\nu}\left(\mathcal{F}_{\rho\sigma}\mathcal{F}_{\tau\kappa}+mB_{\rho\sigma}\mathcal{F}_{\tau\kappa}+\frac{1}{3}m^2B_{\rho\sigma}B_{\tau\kappa}+F^I_{\rho\sigma}F^I_{\tau\kappa}\right)-V_6\right]\,,
\end{align}
where $\tilde{g}$ is the $SU(2)$ gauge coupling constant and $m$ is the mass parameter of the two-form field. The scalar potential is
\begin{equation}
V_6\,=-\frac{1}{8}\left(\tilde{g}^2e^{\sqrt{2}\phi}+4\tilde{g}me^{-\sqrt{2}\phi}-m^2e^{-3\sqrt{2}\phi}\right)\,,
\end{equation}
If we define the superpotential by
\begin{equation}
W_6\,=\,-\frac{1}{4\sqrt{2}}\left(\tilde{g}e^{\frac{\phi}{\sqrt{2}}}+me^{-\frac{3\phi}{\sqrt{2}}}\right)\,,
\end{equation}
then, it gives the scalar potential by
\begin{equation}
V_6\,=\,2\left(\frac{{\partial}W_6}{\partial\phi}\right)^2-5W_6^2\,.
\end{equation}
We employ the mostly-minus signature.

For the supersymmetric $AdS_6$ black hole solutions, we consider the background of
\begin{equation}
ds^2\,=\,e^{2f}\left(dt^2-dr^2\right)-e^{2g_1}\left(d\theta_1^2+\sinh^2\theta_1d\phi_1^2\right)-e^{2g_2}\left(d\theta_2^2+\sinh^2\theta_1d\phi_2^2\right)\,.
\end{equation}
The gauge fields are 
\begin{equation}
A^3\,=\,a_1\cosh\theta_1{d}\phi_1+a_2\cosh\theta_2{d}\phi_2\,,
\end{equation}
where the magnetic charges, $a_1$ and $a_2$, are constant, and the $U(1)$ gauge field is $\mathcal{A_\mu}\,=\,0$.  The two-form field is given by
\begin{equation}
B_{tr}\,=\,-\frac{2}{m^2}a_1a_2e^{\sqrt{2}\phi+2f-2g_1-2g_2}\,,
\end{equation}
and the three-form field strength of the two-form gauge potential vanishes identically. The first order BPS equations are obtained by solving the supersymmetry variations of the fermionic fields,
\begin{align} \label{ads6susy}
f'e^{-f}\,=&\,-\frac{1}{4\sqrt{2}}\left(\tilde{g}e^{\frac{\phi}{\sqrt{2}}}+me^{-\frac{3\phi}{\sqrt{2}}}\right)+\frac{k}{2\sqrt{2}\tilde{g}}e^{-\frac{\phi}{\sqrt{2}}}\left(e^{-2g_1}+e^{-2g_2}\right)-\frac{3}{\sqrt{2}\tilde{g}^2m}e^{\frac{\phi}{\sqrt{2}}-2g_1-2g_2}\,, \notag \\ 
g_1'e^{-f}\,=&\,-\frac{1}{4\sqrt{2}}\left(\tilde{g}e^{\frac{\phi}{\sqrt{2}}}+me^{-\frac{3\phi}{\sqrt{2}}}\right)-\frac{k}{2\sqrt{2}\tilde{g}}e^{-\frac{\phi}{\sqrt{2}}}\left(3e^{-2g_1}-e^{-2g_2}\right)+\frac{1}{\sqrt{2}\tilde{g}^2m}e^{\frac{\phi}{\sqrt{2}}-2g_1-2g_2}\,, \notag \\ 
g_2'e^{-f}\,=&\,-\frac{1}{4\sqrt{2}}\left(\tilde{g}e^{\frac{\phi}{\sqrt{2}}}+me^{-\frac{3\phi}{\sqrt{2}}}\right)-\frac{k}{2\sqrt{2}\tilde{g}}e^{-\frac{\phi}{\sqrt{2}}}\left(3e^{-2g_2}-e^{-2g_1}\right)+\frac{1}{\sqrt{2}\tilde{g}^2m}e^{\frac{\phi}{\sqrt{2}}-2g_1-2g_2}\,, \notag \\ 
\frac{1}{\sqrt{2}}\phi'e^{-f}\,=&\,\frac{1}{4\sqrt{2}}\left(\tilde{g}e^{\frac{\phi}{\sqrt{2}}}-3me^{-\frac{3\phi}{\sqrt{2}}}\right)-\frac{k}{2\sqrt{2}\tilde{g}}e^{-\frac{\phi}{\sqrt{2}}}\left(e^{-2g_1}+e^{-2g_2}\right)-\frac{1}{\sqrt{2}\tilde{g}^2m}e^{\frac{\phi}{\sqrt{2}}-2g_1-2g_2}\,,
\end{align}
and there are twist conditions on the magnetic charges,
\begin{equation}
a_1\,=\,-\frac{k}{\lambda\tilde{g}}\,, \qquad a_2\,=\,-\frac{k}{\lambda\tilde{g}}\,,
\end{equation}
where $k\,=\,\pm1$ are curvatures of the Riemann surfaces and $\lambda\,=\,\pm1$. The $AdS_2\,\times\,\Sigma_{\mathfrak{g}_1}\times\Sigma_{\mathfrak{g}_2}$ horizon solution is given by
\begin{equation} \label{ads6ads2}
e^f\,=\,\frac{2^{1/4}}{g^{3/4}m^{1/4}}\frac{1}{r}\,, \qquad e^{g_1}\,=\,e^{g_2}\,=\,\frac{2^{3/4}}{g^{3/4}m^{1/4}}\,, \qquad e^{\frac{\phi}{\sqrt{2}}}\,=\,\frac{2^{1/4}m^{1/4}}{g^{1/4}}\,,
\end{equation}
where $\mathfrak{g}_1>1$ and $\mathfrak{g}_2>1$ are genus of the Riemann surfaces. The full black hole solutions are interpolating between the $AdS_6$ boundary and the $AdS_2$ horizon.

Now we dimensionally reduce the action on the background of supersymmetric $AdS_6$ black holes. The reduction ansatz for the metric is 
\begin{equation}
ds^2_6\,=\,ds_2^2-e^{2g}\left(ds^2_{\Sigma_{\mathfrak{g}_1}}+ds^2_{\Sigma_{\mathfrak{g}_2}}\right)\,.
\end{equation}
The reduced action is two-dimensional dilaton gravity,
\begin{align}
S\,=\,\frac{vol_{\Sigma_{\mathfrak{g}_1}}vol_{\Sigma_{\mathfrak{g}_2}}}{4\pi{G}_N^{(6)}}\int{d^2x}\sqrt{-g_2}e^{2g_1+2g_2}&\left[-\frac{1}{4}R_2-\frac{1}{2}\partial_\mu{g_1}\partial^\mu{g_1}-\frac{1}{2}\partial_\mu{g_2}\partial^\mu{g_2}-2\partial_\mu{g_1}\partial^\mu{g_2}\right. \notag \\
&\left.+\frac{1}{2}\partial_\mu\phi\partial^\mu\phi-e^{-2g_1-2g_2}V_2\right]\,,
\end{align}
where the scalar potential is
\begin{align}
V_2\,=e^{2g_1+2g_2}&\left[-\frac{1}{8}\left(\tilde{g}^2e^{\sqrt{2}\phi}+4\tilde{g}me^{-\sqrt{2}\phi}-m^2e^{-3\sqrt{2}\phi}\right)-\frac{k_1}{2}e^{-2g_1}-\frac{k_2}{2}e^{-2g_2}\right. \notag \\
&\left.+\frac{1}{2\tilde{g}^2}e^{-\sqrt{2}\phi}\left(e^{-4g_1}+e^{-4g_2}\right)+\frac{2}{\tilde{g}^4m^2}e^{\sqrt{2}\phi-4g_1-4g_2}\right]\,.
\end{align}
The scalar potential satisfies the relations, \eqref{fpoint}, at the $AdS_2$, critical point, \eqref{ads6ads2}. This reduction was previously performed in \cite{Kim:2019fsg}.

We consider the background,
\begin{equation}
ds^2_2\,=\,e^{2f}\left(dt^2-dr^2\right)\,.
\end{equation}
We only consider the case of
\begin{equation}
g\,\equiv\,g_1\,=\,g_2\,>\,1\,, \qquad k\,\equiv\,k_1\,=\,k_2\,=\,-1\,.
\end{equation}
We obtain the flow equations which satisfy the equations of motion of two-dimensional dilaton gravity,
\begin{align}
f'e^{-f}\,=&\,W+\frac{\partial{W}}{\partial{g}}\,, \notag \\
g'e^{-f}\,=&\,W\,, \notag \\
\frac{1}{\sqrt{2}}\phi'e^{-f}\,=&\,-\sqrt{2}\frac{\partial{W}}{\partial\phi}\,,
\end{align}
where the superpotential is given by
\begin{equation}
W=\,-\frac{1}{4\sqrt{2}}\left(\tilde{g}e^{\frac{\phi}{\sqrt{2}}}+me^{-\frac{3\phi}{\sqrt{2}}}\right)-\frac{k}{\sqrt{2}\tilde{g}}e^{-\frac{\phi}{\sqrt{2}}-2g}+\frac{1}{\sqrt{2}\tilde{g}^2m}e^{\frac{\phi}{\sqrt{2}}-4g}\,.
\end{equation}
The superpotential satisfies the relations, \eqref{fpoint2}, at the $AdS_2$ critical point, \eqref{ads6ads2}. The scalar potential is obtained from
\begin{equation}
V_2\,=\,e^{4g}\left[2\left(\frac{\partial{W}}{\partial\phi}\right)^2-2W\frac{\partial{W}}{\partial{g}}-5W^2\right]\,.
\end{equation}
By introducing the superpotential, we note that the flow equations are merely a rewriting of the BPS equations in $AdS_6$, \eqref{ads6susy}, as it should.

Finally, we reproduce the Bekenstein-Hawking entropy of the supersymmetric $AdS_6$ black holes obtained in \cite{Suh:2018tul}, 
\begin{equation}
S_{\text{BH}}\,=\,\left.\frac{e^{4g}vol_{\Sigma_{\mathfrak{g}_1}}vol_{\Sigma_{\mathfrak{g}_2}}}{4G_N^{(6)}}\right|_*\,.
\end{equation}
We make an observation that the two-dimensional dilaton, $\Phi\,\rightarrow\,g$, contains the information of the Bekenstein-Hawking entropy of higher dimensional $AdS$ black holes.

\subsection{Supersymmetric $AdS_5$ black holes}

We review the supersymmetric $AdS_5$ black hole solutions of gauged $\mathcal{N}\,=\,4$ supergravity in five dimensions in \cite{Nieder:2000kc}. See also \cite{Dao:2018xya}. The microstates are microscopically counted by topologically twisted index of $4d$ $\mathcal{N}\,=\,4$ super Yang-Mills theory, \cite{Bae:2019poj}. 

The bosonic field content of $SU(2)\times{U(1)}$-gauged $\mathcal{N}\,=\,4$ supergravity in five dimensions, \cite{Romans:1985ps}, consists of the metric, $g_{\mu\nu}$, a real scalar, $\varphi$,{\footnote{In \cite{Romans:1985ps} the scalar field is parametrized by $\varphi\,=\,\sqrt{\frac{2}{3}}\phi$.}}  an $SU(2)$ gauge field, $A^I_\mu$, $I\,=\,1,\,2,\,3$, a $U(1)$ gauge field, $a_\mu$, and two-form gauge potentials, $B^\alpha_{\mu\nu}$. The field strengths are defined by
\begin{align}
f_{\mu\nu}\,=&\,\partial_\mu{a}_\nu-\partial_\nu{a}_\mu\,, \notag \\
F^I_{\mu\nu}\,=&\,\partial_\mu{A}^I_\nu-\partial_\nu{A}^I_\mu+g_2\epsilon^{IJK}A^J_\mu{A}^K_\nu\,.
\end{align}
The action is given by
\begin{align}
S\,=\,\frac{1}{4\pi{G}_N^{(5)}}\int{d^5x}\sqrt{-g_5}&\left[-\frac{1}{4}R+3\partial_\mu\varphi\partial^\mu\varphi-\frac{1}{4}e^{-4\varphi}f_{\mu\nu}f^{\mu\nu}-\frac{1}{4}e^{2\varphi}\left(F^I_{\mu\nu}F^{I\mu\nu}+B^\alpha_{\mu\nu}B^{\alpha\mu\nu}\right)\right. \notag \\
&\left.+\frac{1}{4}\epsilon^{\mu\nu\rho\sigma\tau}\left(\frac{1}{g_1}\epsilon_{\alpha\beta}B^\alpha_{\mu\nu}D_\rho{B}^\beta_{\sigma\tau}-F^I_{\mu\nu}F^I_{\rho\sigma}a_\tau\right)-V_5\right]\,,
\end{align}
where $g_1$ and $g_2$ are the $U(1)$ and $SU(2)$ gauge coupling constants, respectively, and we define
\begin{equation}
\tilde{g}\,\equiv\,\sqrt{2}g_1\,=\,g_2\,.
\end{equation} 
The scalar potential is
\begin{equation}
V_5\,=-\frac{1}{8}g_2\left(g_2e^{-2\varphi}+2\sqrt{2}g_1e^\varphi\right)\,,
\end{equation}
If we define the superpotential by
\begin{equation}
W_5\,=\,\frac{\tilde{g}}{6\sqrt{2}}\left(2e^{-\varphi}+e^{2\varphi}\right)\,,
\end{equation}
then, it gives the scalar potential by
\begin{equation}
V_5\,=\,\frac{3}{4}\left(\frac{{\partial}W_5}{\partial\varphi}\right)^2-3W_5^2\,.
\end{equation}
We employ the mostly-minus signature.

For the supersymmetric $AdS_5$ black hole solutions, we consider the background of
\begin{equation}
ds^2\,=\,e^{2f}\left(dt^2-dr^2\right)-e^{2g}ds^2_{H_3}\,,
\end{equation}
where
\begin{equation}
ds^2_{H_3}\,=\,d\phi^2+\sinh^2\phi{d}\theta^2+\sinh^2\phi\sin^2\theta{d}\psi^2\,.
\end{equation}
The gauge fields are 
\begin{equation}
A^1_\theta\,=\,a\cosh\phi\,, \qquad A^2_\psi\,=\,b\cos\theta\,, \qquad A^3_\psi\,=\,c\sin\theta\cosh\phi\,,
\end{equation}
where the magnetic charges, $a$, $b$, and $c$, are constant. The $U(1)$ gauge field and two-form field are vanishing. The first order BPS equations are obtained by solving the supersymmetry variations of the fermionic fields,
\begin{align} \label{ads5susy}
f'e^{-f}\,=&\frac{\tilde{g}}{6\sqrt{2}}\left(2e^{-\varphi}+e^{2\varphi}\right)+\frac{\sqrt{2}}{\tilde{g}}e^{\varphi-2g}\,, \notag \\
g'e^{-f}\,=&\frac{\tilde{g}}{6\sqrt{2}}\left(2e^{-\varphi}+e^{2\varphi}\right)-\frac{\sqrt{2}}{\tilde{g}}e^{\varphi-2g}\,, \notag \\
\varphi'e^{-f}\,=&\frac{\tilde{g}}{3\sqrt{2}}\left(e^{-\varphi}-e^{2\varphi}\right)+\frac{\sqrt{2}}{\tilde{g}}e^{\varphi-2g}\,,
\end{align}
and there are twist conditions on the magnetic charges,
\begin{equation}
a\,=\,\frac{1}{\tilde{g}}\,, \qquad b\,=\,\frac{1}{\tilde{g}}\,, \qquad c\,=\,-\frac{1}{\tilde{g}}\,.
\end{equation}
The $AdS_2\,\times\,H_3$ horizon solution is given by
\begin{equation} \label{ads5ads2}
e^{2f}\,=\,\frac{1}{4^{4/3}}\frac{1}{r^2}\,, \qquad e^{2g}\,=\,\frac{1}{4^{1/3}}\,, \qquad e^{3\varphi}\,=\,4\,,
\end{equation}
where $\tilde{g}\,=\,2\sqrt{2}$. The full black hole solutions are interpolating between the $AdS_5$ boundary and the $AdS_2$ horizon.

Now we dimensionally reduce the action on the background of supersymmetric $AdS_5$ black holes. The reduction ansatz for the metric is 
\begin{equation}
ds^2_5\,=\,ds_2^2-e^{2g}ds^2_{H_3}\,.
\end{equation}
The reduced action is two-dimensional dilaton gravity,
\begin{equation}
S\,=\,\frac{vol_{H_3}}{4\pi{G}_N^{(5)}}\int{d^2x}\sqrt{-g_2}e^{3g}\left[-\frac{1}{4}R_2-\frac{3}{2}\partial_\mu{g}\partial^\mu{g}+3\partial_\mu\varphi\partial^\mu\varphi-e^{-3g}V_2\right]\,,
\end{equation}
where the scalar potential is
\begin{equation}
V_2\,=e^{3g}\left[-\frac{1}{8}\tilde{g}^2\left(e^{-2\varphi}+2e^\varphi\right)+\frac{3}{2}e^{-2g}+\frac{3}{2\tilde{g}^2}e^{2\varphi-4g}\right]\,.
\end{equation}
The scalar potential satisfies the relations, \eqref{fpoint}, at the $AdS_2$, critical point, \eqref{ads5ads2}.

We consider the background,
\begin{equation}
ds^2_2\,=\,e^{2f}\left(dt^2-dr^2\right)\,.
\end{equation}
We obtain the flow equations which satisfy the equations of motion of two-dimensional dilaton gravity,
\begin{align}
f'e^{-f}\,=&\,W+\frac{\partial{W}}{\partial{g}}\,, \notag \\
g'e^{-f}\,=&\,W\,, \notag \\
\varphi'e^{-f}\,=&\,-\frac{\partial{W}}{\partial\varphi}\,,
\end{align}
where the superpotential is given by
\begin{equation}
W=\,\frac{\tilde{g}}{6\sqrt{2}}\left(2e^{-\varphi}+e^{2\varphi}\right)-\frac{\sqrt{2}}{\tilde{g}}e^{\varphi-2g}\,.
\end{equation}
The superpotential satisfies the relations, \eqref{fpoint2}, at the $AdS_2$ critical point, \eqref{ads6ads2}. The scalar potential is obtained from
\begin{equation}
V_2\,=\,e^{3g}\left[\frac{3}{4}\left(\frac{\partial{W}}{\partial\varphi}\right)^2-\frac{3}{2}W\frac{\partial{W}}{\partial{g}}-3W^2\right]\,.
\end{equation}
By introducing the superpotential, we note that the flow equations are merely a rewriting of the BPS equations in $AdS_5$, \eqref{ads5susy}, as it should.

Finally, we reproduce the Bekenstein-Hawking entropy of the supersymmetric $AdS_5$ black holes obtained in \cite{Bae:2019poj}, 
\begin{equation}
S_{\text{BH}}\,=\,\left.\frac{e^{3g}vol_{H_3}}{4G_N^{(5)}}\right|_*\,.
\end{equation}
We make an observation that the two-dimensional dilaton, $\Phi\,\rightarrow\,g$, contains the information of the Bekenstein-Hawking entropy of higher dimensional $AdS$ black holes.

\section{Conclusions}

In this paper, we derived the flow equations in two-dimensional dilaton gravity. We commented on holographic $c$-theorem by introducing a quantity which monotonically decreases along the flow. As we have seen in the examples, two-dimensional dilaton gravity is ubiquitous from dimensional reduction of $AdS$ black hole solutions in string and M-theory. Their dynamics and entropy could be understood from two-dimensional perspective.

This opens several intriguing directions we may pursue. Following the studies of higher-dimensional holographic renormalization group flows, \cite{Freedman:1999gp, Skenderis:1999mm, DeWolfe:1999cp}, it is natural to further investigate the flow equations in two-dimensional dilaton gravity. There could be the Hamilton-Jacobi formulation origin of the flow equations analogous to the higher-dimensional flows, \cite{deBoer:1999tgo}. It is also interesting to study the non-perturbative stability of solutions, \cite{Freedman:2003ax}.

There are more supersymmetric black holes in $AdS_4$, \cite{Guarino:2017eag, Guarino:2017pkw, Azzurli:2017kxo, Hosseini:2017fjo, Benini:2017oxt, Bobev:2018uxk}, $AdS_6$, \cite{Suh:2018tul, Hosseini:2018usu, Suh:2018szn}, and $AdS_7$, \cite{Gauntlett:2000ng, Gauntlett:2001jj}. Furthermore, in addition to the magnetically charged static black holes we studied in this work, there is a different class of $AdS$ black holes with electric charges and rotations, $e.g.$, \cite{Hosseini:2017mds, Hosseini:2018dob, Choi:2018fdc}. It would be interesting to consider them from two dimensions.

In relation of $AdS$ black holes from two-dimensional perspective, further understanding of dual $1d$ superconformal quantum mechanics could be pursued, $e.g.$, appendix B of \cite{Benini:2015eyy}.

Our construction provides examples of two-dimensional dilaton gravity from string and M-theory. Although we did not attempt to show by dimensional reduction of higher dimensional gravity, such consistent truncations have been obtained in \cite{Gouteraux:2011qh} for generic dimensions and in \cite{Cvetic:2016eiv} and \cite{Castro:2018ffi} for $AdS_3$ and $AdS_5$ gravity, respectively. See also \cite{Li:2018omr}. It would be interesting to understand the physics of two-dimensional theories along the line of \cite{Almheiri:2014cka, Jensen:2016pah, Maldacena:2016upp, Engelsoy:2016xyb}.

\vspace{1.8cm}

\medskip
\bigskip
\leftline{\bf Acknowledgements}
\noindent We are grateful to Nakwoo Kim and Myungbo Shim for helpful discussions and to Daniel Z. Freedman and Henning Samtleben for instructive communications. Also we would like to thank the anonymous referee for clarifying comments on interpretations of some results. This research was supported by the National Research Foundation of Korea under the grant  NRF-2019R1I1A1A01060811.




\end{document}